\documentclass[reprint,amsmath,amssymb,aps,pra,floatfix,superscriptaddress]{revtex4-2}

\usepackage[T1]{fontenc}
\usepackage{mathptmx}
\DeclareMathAlphabet{\mathcal}{OMS}{cmsy}{m}{n}

\usepackage{graphicx}
\usepackage{dcolumn}
\usepackage{bm}
\usepackage[colorlinks,breaklinks,citecolor=blue,linkcolor=black,urlcolor=black]{hyperref}

\begin{document}

\title{A Gross-Pitaevskii-equation description of the momentum-state lattice: roles of the trap and many-body interactions}%

\author{Tao Chen}
\email{phytch@zju.edu.cn}
\affiliation{Interdisciplinary Center of Quantum Information, State Key Laboratory of Modern Optical Instrumentation, Zhejiang Province Key Laboratory of Quantum Technology and Device, Department of Physics, Zhejiang University, Hangzhou 310027, China}
\affiliation{Department of Physics, University of Illinois at Urbana-Champaign, Urbana, Illinois 61801-3080, USA}
\author{Dizhou Xie}
\affiliation{Interdisciplinary Center of Quantum Information, State Key Laboratory of Modern Optical Instrumentation, Zhejiang Province Key Laboratory of Quantum Technology and Device, Department of Physics, Zhejiang University, Hangzhou 310027, China}

\author{Bryce Gadway}
\email{bgadway@illinois.edu}
\affiliation{Department of Physics, University of Illinois at Urbana-Champaign, Urbana, Illinois 61801-3080, USA}

\author{Bo Yan}
\email{yanbohang@zju.edu.cn}
\affiliation{Interdisciplinary Center of Quantum Information, State Key Laboratory of Modern Optical Instrumentation, Zhejiang Province Key Laboratory of Quantum Technology and Device, Department of Physics, Zhejiang University, Hangzhou 310027, China}
\affiliation{Collaborative Innovation Centre of Advanced Microstructures, Nanjing University, Nanjing, 210093, China}
\affiliation{Key Laboratory of Quantum Optics, Chinese Academy of Sciences, Shanghai, 200800, China}

\date{\today}

\begin{abstract}

We report a theoretical description of the synthetic momentum-state lattices with a 3D Gross-Pitaevskii equation (GPE), where both the external trap potential and the mean-field spatial-density-dependent many-body interactions are naturally included and exactly treated. The GPE models exhibit better performance than the tight-binding model to depict the experimental observations. Since the trap modifies the dispersion relation for free particles and shapes the spatial density distribution that leads to inhomogeneous interactions, decoherences (damping oscillation) appear even for a short-time evolution. Our parametric calculations for the two-state oscillation suggest that we should work with a relatively shallow trap in the weakly interacting regime, especially when the long-term dynamics are concerned. The impact of the mean-field interaction, i.e., the self-trapping behavior, on the transport dynamics and the topological phase transition in a finite multiple-state lattice chain is also specifically investigated. Such an accurate treatment of the inhomogeneous interactions allows for further investigations on the interplay with disorder, the pair correlation dynamics, and the thermalization process in momentum space. 
 
\end{abstract}

\maketitle

\section{introduction}

Synthetic dimensions have opened new avenues for quantum simulations of many interesting and novel phenomena in condensed matter physics \cite{Boada2012, Celi2014}, especially those in connection with topology, disorder, dissipation and Floquet modulation \cite{Ozawa2019, Yuan2018, Goldman2014, Cooper2019}. They also make it feasible to use low-dimensional systems to study new features in higher dimensions \cite{Price2015, Luo2015, Zilberberg2018}. As a recently developed example with ultracold atoms, the momentum-state lattice shows outstanding local control capabilities for all the tight-binding parameters \cite{Gadway2015, Meier2016}: the tunneling strength and phase between each adjacent two sites are set by tuning the Bragg laser intensity and phase of the corresponding frequency component respectively, while the on-site potential is determined by the relative detuning of the Bragg laser pairs. Such precise individual engineering allows easily building lots of models which are typically hard or much more complicated to be realized in real-space optical lattices due to the requirement of flexible separated or time-dependent control. Experimentally, the momentum-state lattice has witnessed success in exploring interesting topological physics, for example, the solitons in the Su-Schrieffer-Heeger (SSH) model \cite{Meier2016a} and an extended variant \cite{Xie2019}, the topological Anderson insulator \cite{Meier2018}, and the topological quantum walk \cite{Xie2020}. By directly adjusting the tunneling phases to form an artificial magnetic field, the edge current and the non-reciprocal transport have also been observed in momentum-state flux lattice \cite{An2017, Gou2020}. Additionally, the Anderson localization and the mobility edge have been carefully investigated by introducing programmed disorder and variations to the on-site potentials \cite{An2018a, An2021}.

Generally, the above mentioned phenomena realized in momentum-state lattice can be well described by the single-particle tight-binding model \cite{Gadway2015}. The derivation of the tight-binding Hamiltonian there starts from the light-matter interaction theory, where the effects from both the relatively weak optical trapping and atomic interactions are neglected. Whereas such approximations are safe as our experimental observations are highly consistent with the theoretical predictions, a more precise description is still lacking and eagerly required, especially when we go beyond the weakly trapping and interacting regime, for example, with the Feshbach resonance technique to tune the $s$-wave scattering length \cite{Chin2010}. Qualitatively speaking, the external trap modifies the dispersion relation and shapes the atomic density distribution, while the interactions leads to a density-dependent attractive on-site potential \cite{An2018}. Although the self-trapping feature can be resolved by directly adding an averaged mean-field interaction terms to the single-particle Sch\"odinger equation in momentum space \cite{An2018, Xie2020}, the information of the inhomogeneous spatial distribution (thus the inhomogeneous interaction strength) of the atomic cloud is to a large extent missing and inaccurately accounted.

In this work, we develop a GPE description to the momentum-state lattice, where naturally the external trapping and the many-body interaction effect are taken into consideration. We find that such a GPE treatment can well interpret the experimentally observed two-level Rabi oscillation with decoherences. We also investigate the effect of the interactions by artificially changing the total atomic number and the $s$-wave scattering length. The dependences of the oscillation parameters on the trap frequencies and the interaction strength are presented. The paper is organized as follows. Section \ref{sec2} summarizes the derivation of the GPE under the Bragg laser field. In Sec.\ref{sec3}, we show the parameter dependences and the difference between the tight-binding model and the GPE approach. In Sec.\ref{sec4}, we give two case studies to show the effect of the interaction induced localization on the state transport and topological phase transition in a finite-length chain. Section \ref{sec5} concludes and gives an outlook.

\section{The GPE model}\label{sec2}

\begin{figure}[b]
\includegraphics[width=0.45\textwidth]{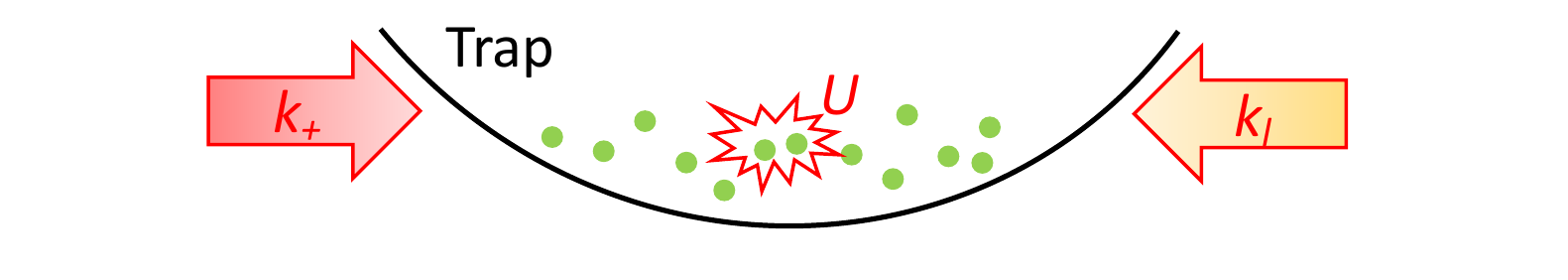}
\caption{\label{fig0} Schematic illustration of the momentum-state lattice. A trapped condensate interacts with a counter-propagating Bragg laser pair along $z$-direction. The interaction strength of the atoms is $U$. The right beam has multiple frequency components to respectively couple different momentum-state pairs.
}
\end{figure}

The momentum-state lattice is built with a weakly trapped Bose-Einstein condensate interacting with the Bragg laser pairs, as shown in Fig.~\ref{fig0}. Assuming the Bragg laser beams propagate along the $z$-direction, the field is described by
\begin{eqnarray}
\mathcal{E} &=& \frac{1}{2}\hat{\epsilon}_+ e^{i(k_+z-\omega_+t + \tilde{\phi}_+)} + \sum\limits_\ell \frac{1}{2}\hat{\epsilon}_\ell e^{i(-k_\ell z-\omega_\ell t + \tilde{\phi}_\ell)} + {\rm c.c.},
\end{eqnarray}
where $k_+\simeq k_\ell = k$. By treating the atom as a two-level system ($|g\rangle$ and $|e\rangle$, the transition frequency is $\omega_{\rm eg}$), under the far-off-resonance condition (i.e., the detuning $\Delta=\omega_+ - \omega_{\rm eg}$ is much larger than the linewidth of the excited state) and the rotating wave approximation, we can adiabatically eliminate the excited state $|e\rangle$ and obtain the time-dependent optical potential of the Bragg lasers as \cite{Blakie2000, Potting2001}
\begin{equation}
V_{\rm Bragg}(z, t) = \sum\limits_\ell 2\hbar J_\ell \cos (2kz - \delta_\ell t + \phi_\ell),
\end{equation}
where $\delta_\ell = \omega_+ - \omega_\ell$, $\phi_\ell = \tilde{\phi}_+ - \tilde{\phi}_\ell$ and $J_\ell = \Omega_+\Omega_\ell/4|\Delta|$ with $\Omega_{+(\ell)} = -\langle e| \hat{d}\cdot \hat{\epsilon}_{+(\ell)}|g\rangle/\hbar$ (here $\hat{d}$ is the dipole moment of the atom). Then, the time evolution of the mean-field wavefunction of the condensate, $\Psi (\mathbf{r}, t)$, follows the Gross-Pitaevskii equation \cite{Dalfovo1999}
\begin{equation}\label{eq3}
i\hbar\frac{\partial\Psi}{\partial t} = \left(-\frac{\hbar^2\nabla^2}{2\mu} + V_{\rm trap} + V_{\rm Bragg} + U|\Psi|^2\right)\Psi, 
\end{equation}
where the trap potential $V_{\rm trap} = \frac{1}{2}\mu(\omega_x^2 x^2 + \omega_y^2 y^2 + \omega_z^2 z^2 )$ with $(\omega_x, \omega_y, \omega_z)$ the trap frequencies, and the interaction strength $U = 4\pi\hbar^2 (N-1)a_s/\mu$ with $N$ the atom number and $a_s$ the $s$-wave scattering length. Here $\Psi (\mathbf{r}, t)$ fulfills the normalization $\int d\mathbf{r} |\Psi(\mathbf{r},t)|^2 = 1$.

To clearly illustrate the periodic lattice structure in momentum space, we decompose the mean-field wavefunction $\Psi (\mathbf{r},t)$ with the ansatz \cite{Wu2000, Gadway2015}
\begin{equation}\label{eq4}
 \Psi(\mathbf{r},t) = \sum\limits_n \psi_n(\mathbf{r},t) e^{2i nkz},
\end{equation}
with $n\in \mathbb{Z}$. Such a decomposition validates according to the fact that $\psi_n(\mathbf{r},t)$ is localized with a rather narrow broadening at $p_n=2n\hbar k$  in momentum space, as demonstrated in the experiment (also in our later calculation in Sec.~\ref{sec3}). Following the standard mean-field approach \cite{Dalfovo1999}, we obtain the coupled time-dependent GPEs for each $\psi_n(\mathbf{r},t)$ by making a variation to the energy functional, $E = \int d\mathbf{r} [\frac{\hbar^2}{2\mu}|\nabla\Psi|^2 + (V_{\rm trap}+V_{\rm Bragg})|\Psi|^2 + \frac{U}{2}|\Psi|^4]$, with respect to $\psi_n^*(\mathbf{r},t)$, as following
\begin{eqnarray}\label{eq5}
 i\hbar\frac{\partial \psi_n}{\partial t} &=& \left(-\frac{\hbar^2\nabla^2}{2\mu} + 4n^2E_r - \frac{2n\hbar k}{\mu} i\hbar\partial_z + V_{\rm trap}\right)\psi_n \nonumber\\[0.5em]
 &+& \sum\limits_\ell \left( J_\ell e^{-i\phi_\ell}e^{i\delta_\ell t}\psi_{n+1} + J_\ell e^{i\phi_\ell}e^{-i\delta_\ell t}\psi_{n-1}\right) \nonumber\\[0.5em]
 &+& U(|\psi_n|^2+\sum_{j\neq n}2|\psi_j|^2 )\psi_n,
\end{eqnarray}
where $E_r = \hbar^2k^2/(2\mu)$ is the photon recoil energy. Under frame rotation, we let $\varphi_n (\mathbf{r},t)= e^{i4n^2E_rt/\hbar}\psi_n(\mathbf{r},t)$, then Eq.(\ref{eq5}) reduces to
\begin{eqnarray}\label{eq6}
 i\hbar\frac{\partial \varphi_n}{\partial t} &=& \left(-\frac{\hbar^2\nabla^2}{2\mu} - \frac{2n\hbar k}{\mu} i\hbar\partial_z + V_{\rm trap} \right)\varphi_n \nonumber\\[0.5em]
 &+& \sum\limits_\ell J_\ell e^{-i\phi_\ell}e^{i[\delta_\ell-4(2n+1)E_r/\hbar] t}\varphi_{n+1} \nonumber\\[0.5em]
 &+& \sum\limits_\ell J_\ell e^{i\phi_\ell}e^{-i[\delta_\ell -4(2n-1)E_r/\hbar]t}\varphi_{n-1} \nonumber\\[0.5em]
 &+& U(|\varphi_n|^2+\sum_{j\neq n}2|\varphi_j|^2)\varphi_n.
\end{eqnarray}
The second and third terms in the right hand of Eq.(\ref{eq6}) take into account all the frequency components which contribute to $|p_{n \pm 1}\rangle\leftrightarrow |p_n\rangle$ transition. If we simply choose the detuning $\delta_\ell = 4(2\ell+1)E_r/\hbar$, and neglect the off-resonant terms, the above equation can be written as
\begin{eqnarray}\label{eq7}
 i\hbar\frac{\partial \varphi_n}{\partial t} &=& \left(-\frac{\hbar^2\nabla^2}{2\mu} - \frac{2n\hbar k}{\mu} i\hbar\partial_z + V_{\rm trap} \right)\varphi_n \nonumber\\[0.5em]
 &+ & J_n e^{-i\phi_n}\varphi_{n+1} + J_{n-1}e^{i\phi_{n-1}}\varphi_{n-1} \nonumber\\[0.5em]
 &+ & U(|\varphi_n|^2+\sum_{j\neq n}2|\varphi_j|^2)\varphi_n.
\end{eqnarray}
This coupled GPE formula can be directly mapped into the tight-binding model in momentum space \cite{Gadway2015}, but with additional information of the trap and the spatial-dependent interaction involved. The $i\hbar\partial_z$ term indicates the spatial expansion of the wavefunction $\varphi_n$ along the $z$-direction, i.e., the density flow of the condensate as the atoms have a momentum of $p_n$. For long-time evolutions, this leads to the atoms with large $p_n$ rapidly reach the boundary of the trap and consequently hinders the coherent transport dynamics. We also note that, here $\varphi_n$ is spatial dependent, and consequently the interaction term differs from the form in the nonlinear Sch\"odinger equation in Ref.~\cite{An2018}, where a uniform density distribution is assumed to arrive at an equivalent attractive density-dependent on-site potential. In the following calculations, we numerically solve Eq.(\ref{eq3}) or Eq.(\ref{eq7}) to study the time evolution of the initial condensate in momentum space, where the trap frequencies and the interaction strength indeed play essential roles. 

\begin{figure}[]
\includegraphics[width=0.5\textwidth]{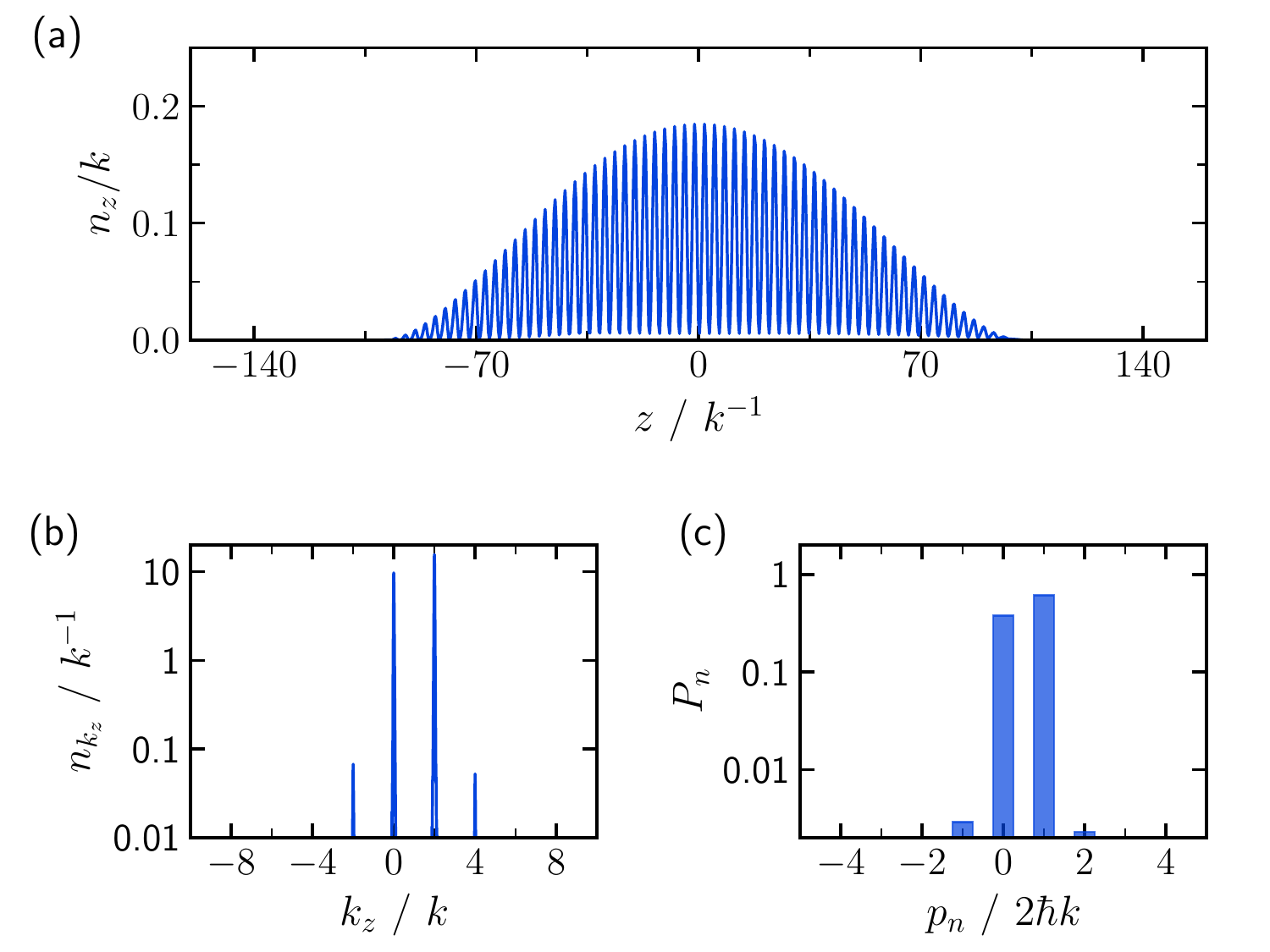}
\caption{\label{fig1} Time evolution of a $^{87}$Rb condensate under a pair of Bragg lasers (wavelength $\sim 1064~{\rm nm}$) with $\ell = 0$ and $\delta_\ell = 4E_r/\hbar$.
(a) Density distribution $n_z$ along the $z$-direction after the condensate evolves a time of $t=0.16 ~{\rm ms}$ with Eq.(\ref{eq3}).
(b) Density distribution $n_{k_z}$ along the $k_z$ momentum space, which is calculated by making a 3D Fourier transformation to the wavefunction $\Psi(\mathbf{r}, t=0.16~{\rm ms})$.
(c) Populated fraction $P_{n}$ for each discrete momentum state $|p_n\rangle$ from respective integrations of each peak in (b).
During the evolution, we use parameters: $N=1\times 10^5$, $a_s = 100 a_0$ ($a_0$ is Bohr radius), $\omega_x = \omega_y = 2\pi \times 200~{\rm Hz}$, $\omega_z=2\pi \times 40~{\rm Hz}$, $J_{\ell=0} = 2\pi \times 1000~{\rm Hz}$.
}
\end{figure}

\section{two-state oscillation}\label{sec3}

\subsection{Fully exact treatment}

We first study the simplest case, i.e., the oscillation between two momentum states, by directly evolving the GPE (\ref{eq3}) with $\ell = 0$ and $\delta_\ell = 4E_r/\hbar$. We note that, similar works on the double-well oscillating behavior and swallowtail-type state transfer under interactions have been reported \cite{Guan2020, Guan2020a}. Here we just outline the details on how to obtain the atomic fraction in each momentum state with a 3D Fourier transformation. Initially, the wavefunction $\Psi(\mathbf{r},t=0)$ is obtained by solving the ground state of the GPE without the Bragg laser field. After the Bragg laser is applied for $t=0.16~{\rm ms}$, the spatial density distribution along $z$-direction, $n_z$, significantly changes into a form with multiple fringe peaks, as shown in Fig.~\ref{fig1}(a). $n_z(z,t)$ is calculated from the mean-field wavefunction $\Psi(\mathbf{r},t)$ as
\begin{equation}
 n_z(z,t) = \int\int dx dy |\Psi(\mathbf{r},t)|^2.
\end{equation}

To yield the atom number in each momentum state, we make a 3D Fourier transformation to $\Psi(\mathbf{r},t)$ to obtain the momentum-space wavefunction $\Psi(\mathbf{k},t)$. Then, the density distribution along the $k_z$-direction can be calculated with 
\begin{equation}\label{eq9}
 n_{k_z}(k_z,t) = \int\int dk_x dk_y |\Psi(\mathbf{k},t)|^2.
\end{equation}
Figure \ref{fig1}(b) shows an example of $n_{k_z}$ corresponding to the spatial density distribution in Fig.~\ref{fig1}(a). Four narrow peaks are resolved, which provides an additional evidence to validate the decomposition (\ref{eq4}). While the Bragg laser pair only resonantly couples momentum states $|p_0 = 0\rangle$ and $|p_1 = 2\hbar k\rangle$, we observe the non-resonant couplings (with detunings of $\sim 8E_r$) induced slight population leakage into $|p_{-1}\rangle$ and $|p_2\rangle$. Finally, as shown in Fig.~\ref{fig1}(c), we resolve the population fraction $P_n$ for the momentum state $|p_n = 2n\hbar k\rangle$ by integrating each peak along $k_z$ as
\begin{equation}
 P_n(t) = \int_{(2n-1)k}^{(2n+1)k} dk_z~ n_{k_z}(k_z,t).
\end{equation}
This is the general procedure via which we obtain the dynamics of the condensate in momentum space with Eq.~(\ref{eq3}). Here we take a two-state oscillation as an example, whereas it can be extended to multi-state cases by simply adding frequency components to the Bragg lasers. 

\subsection{Validation of the approximations}

Here we focus on the time evolution of the condensate in momentum space, thus only the population oscillation between each momentum state is concerned. We compare the oscillation dynamics obtained with Eq.(\ref{eq3}), Eq.(\ref{eq6}) and Eq.(\ref{eq7}) respectively for a two-state case to show the validation of both the decomposition of the mean-field wavefunction and the elimination of the non-resonant coupling terms. During the propagations of Eqs.(\ref{eq6}) and (\ref{eq7}), the population fraction in $|p_n\rangle$ is obtained with $P_n(t) = \int d\mathbf{r} |\varphi_n(\mathbf{r},t)|^2$.

\begin{figure}[]
\includegraphics[width=0.5\textwidth]{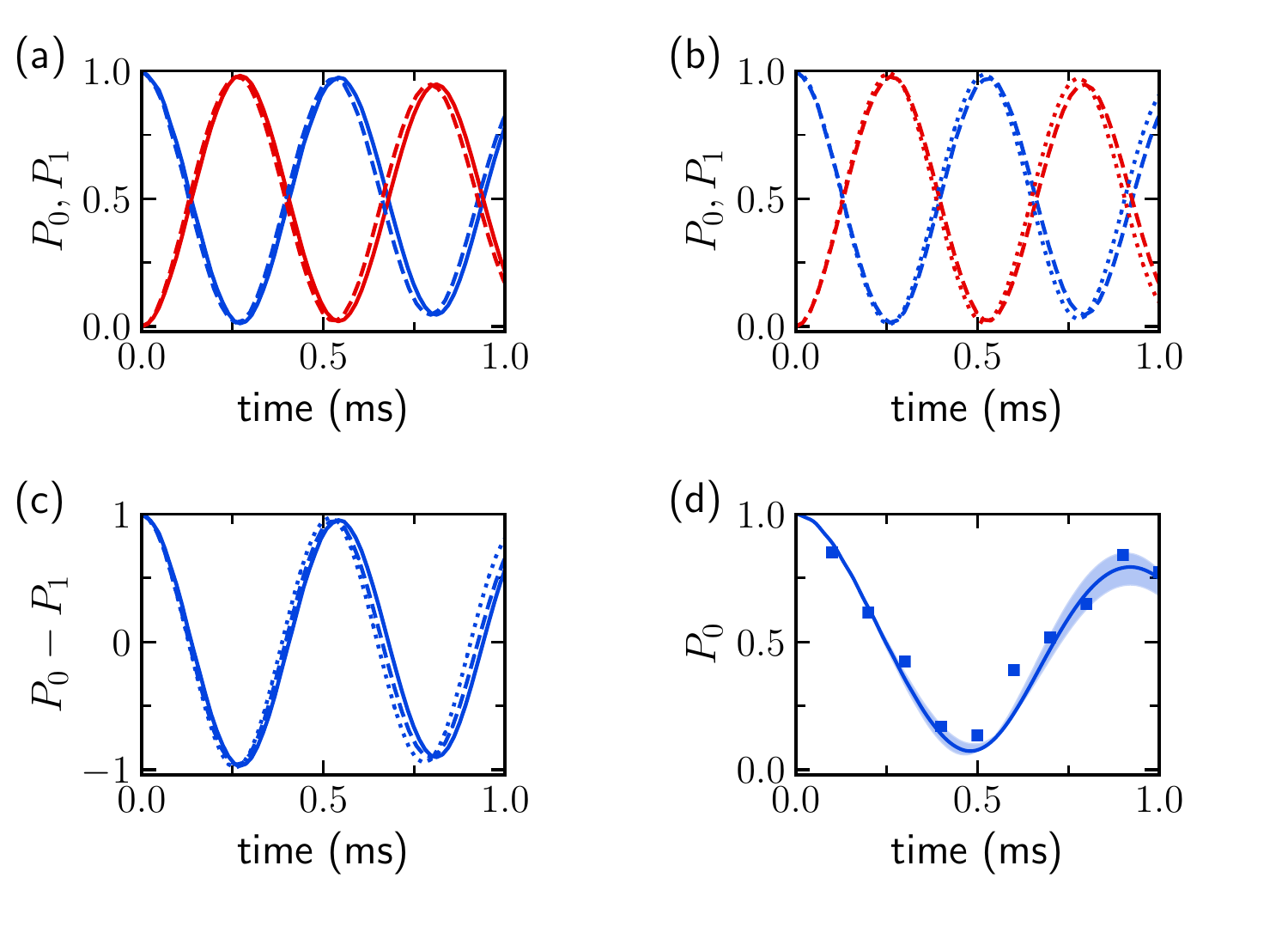}
\caption{\label{fig2} (Color online) (a)-(c) Comparisons of the dynamics derived from exact and approximated treatments. Two-state oscillations and the time evolution of the population difference $P_0-P_1$are shown respectively. During the evolution, we use parameters: $N=1\times 10^5$, $a_s = 100 a_0$, $\omega_x = \omega_y = 2\pi \times 200~{\rm Hz}$, $\omega_z=2\pi \times 40~{\rm Hz}$, $J_{\ell=0} = 2\pi \times 1000~{\rm Hz}$. Here the results from Eq.(\ref{eq3}), (\ref{eq6}) and (\ref{eq7}) are marked with solid, dashed and dotted lines respectively. In (a) and (b), $P_0$ and $P_1$ are indicated by red and blue colors respectively. (d) Comparison between the experimental measured dynamics (square points) with the GPE result. The parameters in our experiment are: $N=1.1(3)\times 10^5$, $J_{\ell=0}=2\pi\times 600(20)~{\rm Hz}$, $(\omega_x, \omega_y, \omega_z)=2\pi\times (175, 175, 60)~{\rm Hz}$. The solid line is the numerically calculated result with Eq.(\ref{eq3}) under experimental parameters. The shaded region results from the upper and lower bounds of the parameters.
}
\end{figure}

Figure \ref{fig2}(a) shows the oscillations between $|p_0\rangle$ and $|p_1\rangle$ calculated with Eq.(\ref{eq3}) and Eq.(\ref{eq6}). Obviously, the two results coincide with each other, especially for short evolution time ($t< 0.5 ~{\rm ms}$) with the parameters there. This indicates that the decomposition (\ref{eq4}) of the mean-field wavefunction $\Psi(\mathbf{r},t)$ is at least a good approximation when only the population dynamics are concerned. On the other hand, we compare the result from Eq.(\ref{eq7}) by dropping the non-resonant terms with that from Eq.(\ref{eq6}). As shown in Fig.~\ref{fig2}(b), the behaviors of the two dynamics are again consistent with each other, as the tunneling amplitude $J_{\ell=0} \ll 8E_r$ here. The effect of the non-resonant tunneling terms is indicated by the relatively larger amplitude of the oscillation from Eq.(\ref{eq7}) (dotted lines), since they contribute to leakage to $|p_{-1}\rangle$ and $|p_2\rangle$ states, which degrades the populations in $|p_0\rangle$ and $|p_1\rangle$. From the comparisons, we conclude that the approximated coupled GPEs (\ref{eq7}) can effectively describe the dynamics of the momentum-state lattice, for a finite evolution time under an appropriate Bragg coupling strength.

Different from the ideal two-level Rabi oscillation resolved from the tight-binding model, the dynamics calculated here exhibit enlarged oscillating periods and decoherences, as shown in Fig.~\ref{fig2}(c). For an ideal two-state Rabi oscillation with a tunneling $J = 2\pi \times 1000~{\rm Hz}$, a $2\pi$-pulse corresponds to an evolution time of exactly $0.5~{\rm ms}$. However, in Fig.~\ref{fig2}(c), it takes a time interval larger than $0.5~{\rm ms}$ to let the atoms populate back to $|p_0\rangle$. More importantly, the atoms can not all return back due to the slightly self-trapping. As such phenomena still exist even when working with Eq.(\ref{eq7}), we attribute the enlarged period and decoherence to the effect of the weak trap potential and the interaction terms in the GPEs.

Besides the approximations from Eq.(\ref{eq3}) to Eq.(\ref{eq7}), we demonstrate the validation of the approximation to yield the Bragg laser potential by making a comparison between the experimental measured oscillation and that calculated with Eq.(\ref{eq3}), as shown in Fig.~\ref{fig2}(d).

\subsection{Parameter dependence of the oscillation}

\begin{figure}[]
\includegraphics[width=0.5\textwidth]{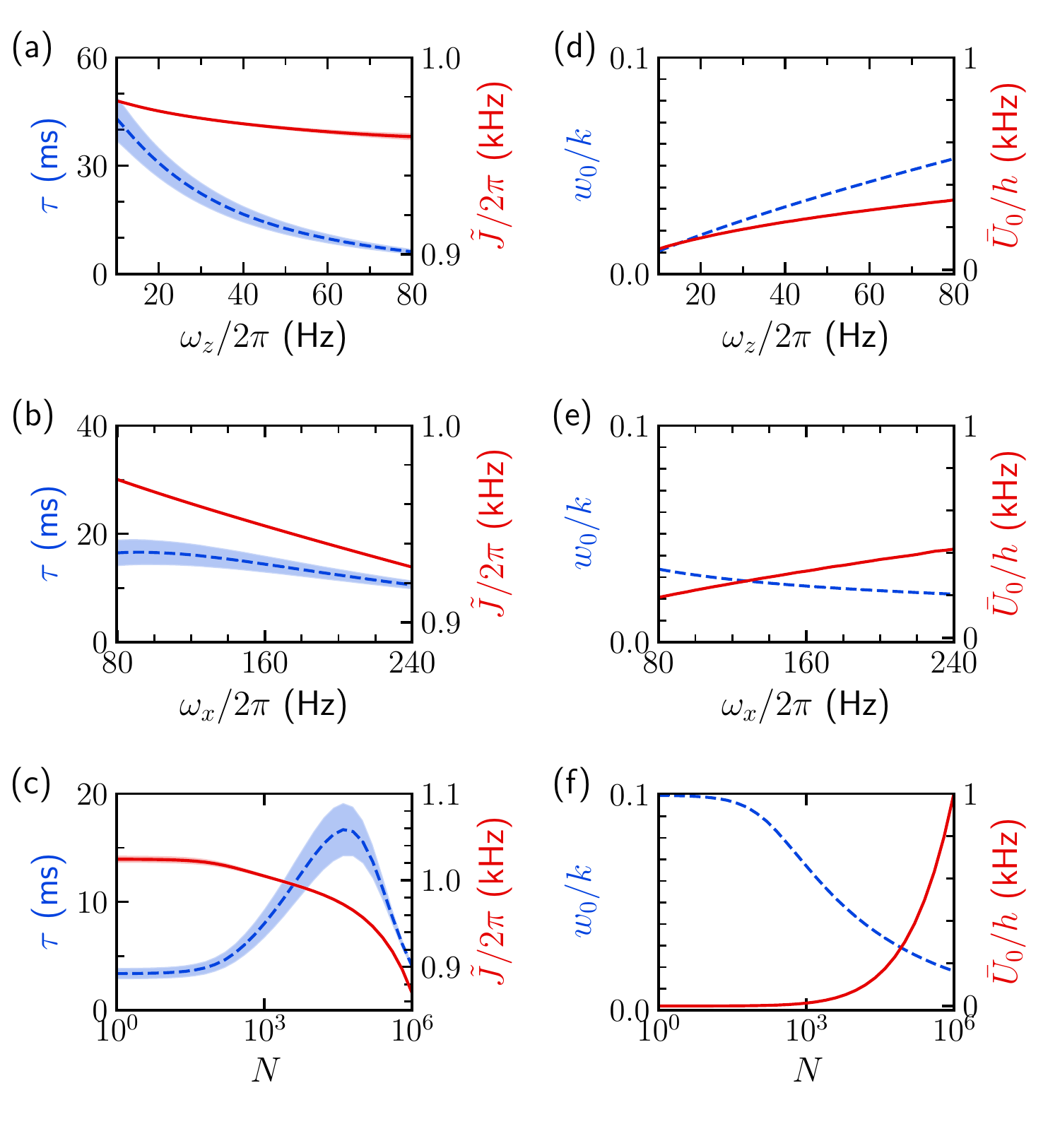}
\caption{\label{fig3} (Color online) (a)-(c) Dependences of $\tau$ (blue dashed lines) and $\tilde{J}$ (red solid lines) on: the trap frequencies $\omega_z$ (a) and $\omega_x$ (b), and the total atom number $N$ (c). (d)-(f) Dependences of $w_0$ (blue dashed lines) and $\bar{U}_0$ (red solid lines) on: the trap frequencies $\omega_z$ (d) and $\omega_x$ (e), and the total atom number $N$ (f). Here the scattering length $a_s = 100 a_0$, and we fix $\omega_x = \omega_y$, $J_{\ell=0} = 2\pi \times 1000~{\rm Hz}$. In (a) and (d), $N = 6\times 10^4$ and $\omega_x = 2\pi \times 100~{\rm Hz}$, while in (b) and (e), $N = 6\times 10^4$ and $\omega_z = 2\pi\times 40~{\rm Hz}$. In (c) and (f), $\omega_x = 2\pi \times 100~{\rm Hz}$ and $\omega_z = 2\pi\times 40~{\rm Hz}$. The shaded regions in (a)-(c) indicate 0.95-confidence-intervals from the fittings with Eq.~(\ref{eq11}).
}
\end{figure}

We parameterize the two-state damping oscillation in Fig.~\ref{fig2}(c) with two variables: the decoherence time constant $\tau$, and the actual tunneling rate $\tilde{J}$, which can be resolved by fitting the calculated $P_0-P_1$ result with a function
\begin{equation}\label{eq11}
 f(t) = e^{-t/\tau} \cos(2\tilde{J}t).
\end{equation}
In the following, we make detailed investigations on the dependences of both $\tau$ and $\tilde{J}$ on the trap frequencies and the interaction strength. Throughout this section, we directly propagate the GPE (\ref{eq3}) rather than the approximated one (\ref{eq7}) to yield the dynamics within $0.8~{\rm ms}$. The effects of the trap potential and interactions are respectively monitored by the initial momentum broadening width $w_0$ and the initial spatially averaged interaction strength $\bar{U}_0$. The broadening width $w_0$ is obtained from a gauss fit to the initial density distribution along the $k_z$-direction, $n_{k_z}(t=0)$ [see Eq.(\ref{eq9})]; while
\begin{equation}\label{eq12}
 \bar{U}_0= \frac{1}{\mathcal{V}}\int d\mathbf{r} U |\Psi(\mathbf{r}, t=0)|^2,
\end{equation}
with the volume $\mathcal{V}=\int_{|\Psi(\mathbf{r}, t=0)| \ge 10^{-3}} d\mathbf{r}$.

Figures \ref{fig3}(a) and (b) show how the oscillation parameters vary versus the trap frequencies. Generally, both $\tau$ and $\tilde{J}$ decrease when the trap becomes tighter, either along $z$-direction or the radial $x(y)$-direction. We note that $\tau$ is much more sensitive to $\omega_z$ than to $\omega_x$, as the value of $\omega_z$ dominates the initial momentum broadening of the condensate along $k_z$ [see Figs.~\ref{fig3}(d) and (e)]. Not surprisingly, the fitted actual tunneling rates $\tilde{J}$ are all smaller than used $J$. This leads to a reduction of the tunneling by less than 10\% for typical trap frequencies. Here we should emphasize that, the behaviors of $\tau$ and $\tilde{J}$ in Figs.~\ref{fig3}(a) and (b) are caused by both trap potential and the interactions, although only the trap frequencies get changed. In fact, for fixed atom number and $s$-wave scattering length, a tighter trap leads to a higher spatial density which results in relatively stronger many-body interactions; see $\bar{U}_0$ shown in Figs.~\ref{fig3}(d) and (e). 

The effects from the trap potential and the interactions compete with each other. As shown in Fig.~\ref{fig3}(e), by increasing $\omega_x$, the broadening width $w_0$ slightly decreases, which should make the decoherence constant $\tau$ slightly increase. However, due to the increase of the interaction strength, we still observe a decrease of $\tau$ in Fig.~\ref{fig3}(b). We also note that, the values of $\tilde{J}$ are much more likely determined by the interaction strength, but not sensitive to the initial momentum broadening.  

To directly illustrate the effect of the interactions, we plot the fitted $\tau$ and $\tilde{J}$ as functions of the total atom number $N$ in Fig.~\ref{fig3}(c). The competition between the trap potential and the interactions is clearly verified, as we find a critical value of $N$ where the decoherence constant $\tau$ reaches its maximum. This can be easily understood from Fig.~\ref{fig3}(f): for small $N$, the effect of the interactions can be neglected, the initial momentum broadening determines the tendency of $\tau$; while for large $N$, the strong interaction strength plays a dominant role and leads to a rapid decrease of both $\tau$ and $\tilde{J}$. 

\begin{figure}[]
\includegraphics[width=0.5\textwidth]{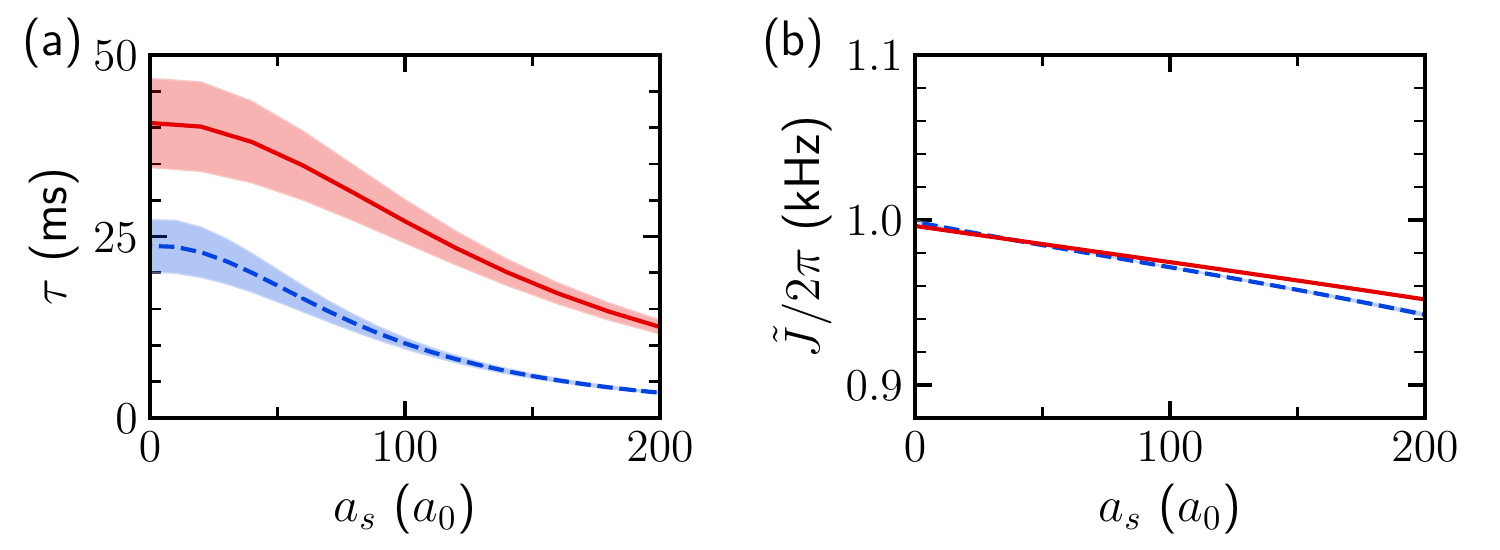}
\caption{\label{fig4} (Color online) Dependences of $\tau$ (a) and $\tilde{J}$ (b) on the tunable $s$-wave scattering length $a_s$ for $t>0$. Here the condensate is prepared ($t<0$) with a scattering length $a_s = 100 a_0$, and the trap frequencies $\omega_x = \omega_y = 2\pi \times 100~{\rm Hz}$, and the atom number $N=6\times 10^4$. At $t=0$, we turn on the Bragg laser with $J = 2\pi\times 1000~{\rm Hz}$, let the trap off and change the scattering length to a desired value. The blue dashed lines and red solid lines correspond to $\omega_z=2\pi\times 40~{\rm Hz}$ and $\omega_z=2\pi\times 20~{\rm Hz}$ repectively. The shaded regions indicate 0.95-confidence-intervals from the fittings.
}
\end{figure}

To better establish the effect purely from the many-body interactions, we perform calculations with an assumed tunable $s$-wave scattering length. Initially ($t<0$), we prepare a condensate with a fixed atom number $N$ and $a_s=100 a_0$ in the trap. At $t=0$, the Bragg laser is applied and simultaneously we turn off the trap and tune the scattering length. Now both the initial momentum broadening $w_0$ and the initial shape are all identical to each other respectively, the only difference for the evolutions in $t>0$ is the interaction strength with different $a_s$ values, as no trap exists yet. Figure \ref{fig4} illustrates the dependences of $\tau$ and $\tilde{J}$ on the scattering length $a_s$ ($t>0$). The maximum value of decoherence constant $\tau$ shows up at $a_s=0$, where the influences on the two-state oscillation from both the trap potential and the interaction are eliminated. The finite value of $\tau$ at $a_s=0$, in comparison with the infinite value ($\tau\to\infty$) for an ideal two-level Rabi oscillation, results from the initial momentum broadening $w_0$ which rapidly decreases and hence leads to larger $\tau$ when we work with a weaker trap potential along the $z$-direction [see Fig. \ref{fig3}(d)]. When we enlarge the scattering length, $\tau$ rapidly decreases as the interaction strength that is proportional to $a_s$ become stronger. 

\section{Multiple-state transport}\label{sec4}

\subsection{Self-trapping in a uniform chain}

\begin{figure}[]
\includegraphics[width=0.5\textwidth]{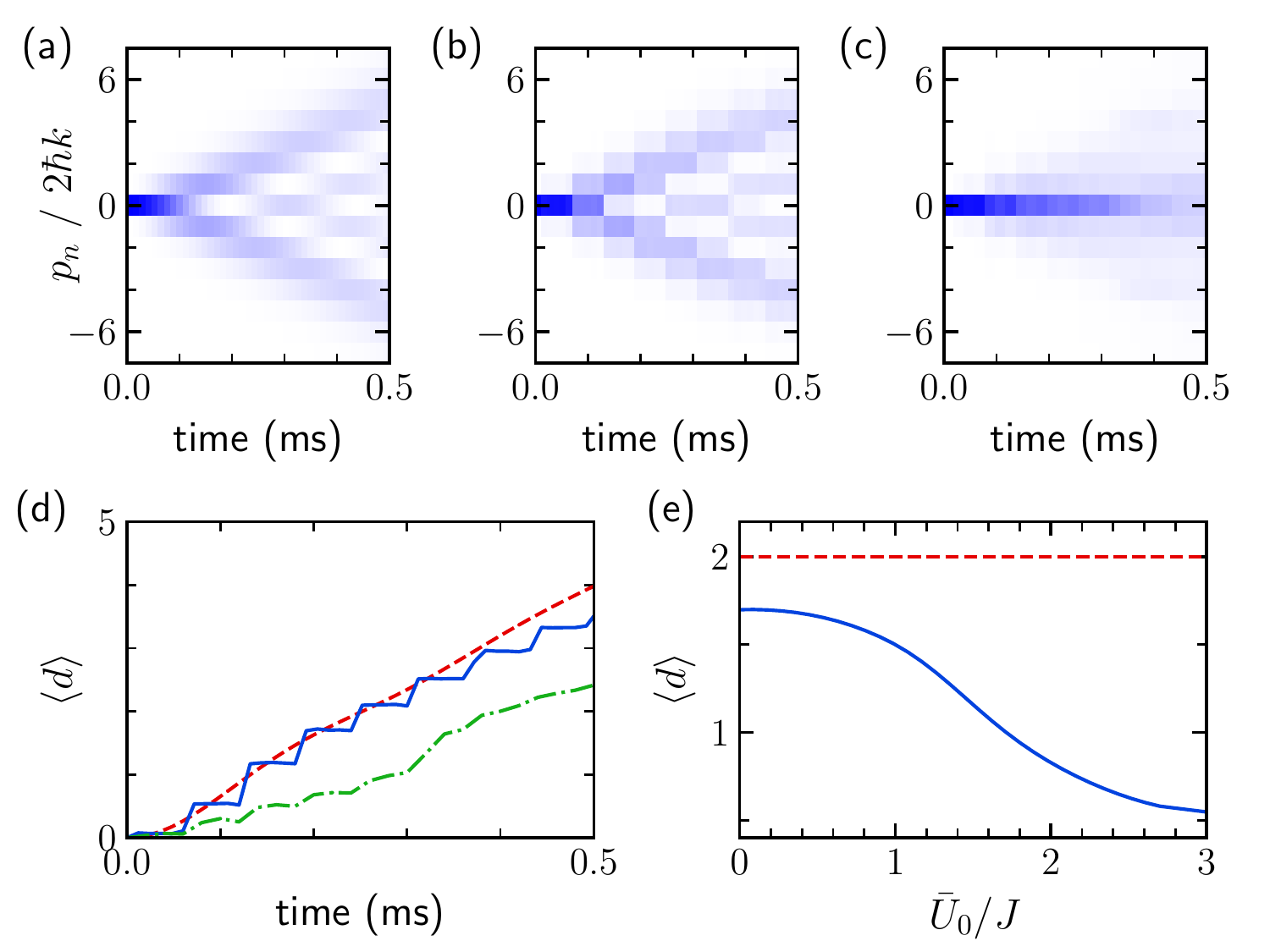}
\caption{\label{fig7} (Color online) Interaction induced localization in a 15-site uniform momentum-state lattice chain. Time evolution of the populations in each site are  obtained from: (a) a tight-binding Hamiltonian $H = \sum_\ell J^{}_\ell c_{\ell+1}^\dagger c_\ell^{} + {\rm h.c.}$, or the GPE model (\ref{eq3}) with (b) $a_s = 100 a_0$ and (c) $a_s=1000 a_0$ for $t>0$. Here we choose $J_\ell = 2\pi\times 1000~{\rm Hz}$. In (b) and (c), the condensate ($N=6\times 10^4$) is prepared ($t<0$) in a trap with frequencies: $\omega_x=\omega_y=2\pi\times 100~{\rm Hz}$, $\omega_z=2\pi\times 40~{\rm Hz}$, and the scattering length $a_s=100 a_0$. (d) Evolution of the mean propagated distance $\langle d\rangle$ of the wavepacket in momentum space. The dashed red line, the solid blue line and the dash dotted line correspond to the cases in (a)-(c) respectively. (e) Dependence of $\langle d\rangle$ at time $t=0.24~{\rm ms}$ on the interaction strength $\bar{U}_0$. Since $\bar{U}_0\propto a_s(t=0)$, here we simply fix $J_\ell = J = 2\pi\times 1000~{\rm Hz}$ while changing $a_s$ to vary the ratio $\bar{U}_0/J$. The dashed red line indicates the ideal limit calculated with the tight-binding Hamiltonian. 
}
\end{figure}

The interaction induced localization, i.e., the so-called self-trapping in momentum-state lattices has been both theoretically and experimentally investigated \cite{An2018, Xie2020}. In previous studies, the inhomogeneous density-dependent interactions are treated with a weighted average to the dynamics obtained from a series of nonlinear Sch\"odinger equations with different interaction strength varying from zero to a peak value. While such an average can well describe the experimental observations, our GPE approach is a more precise model as it takes into account the spatial density distribution which dynamically evolves under the Bragg laser field. 

The self-trapping behavior is presented by comparing the dynamics with interactions and that for an ideal case. As shown in Fig.~\ref{fig7}(a), the condensate undergoes a quantum walk in momentum space before reflected back by the boundary when the tight-binding Hamiltonian is applied for a 15-site uniformly coupled chain. Under weak interactions, the transport along the chain still effectively happens; see Fig.~\ref{fig7}(b) with $a_s=100a_0$. When the interaction becomes strong, for example, $a_s=1000a_0$ in Fig.~\ref{fig7}(c), the atoms tend to be localized in the central several sites. As revealed in Ref.~\cite{An2018}, this localization behavior is unique for atomic momentum-state lattice under many-body interactions which introduce attractive on-site potentials. A strong interaction strength breaks (weakens) the tunneling processes as the attractive potentials can be regarded as energy shifts which make the Bragg lasers far off-resonance. 

To quantitatively describe the transition from diffusion to localization, we use the mean propagated distance along the momentum space, $\langle d\rangle = \sum_n |n|P_n$, to indicate the transport ability. As shown in Fig.~\ref{fig7}(d), $\langle d\rangle$ linearly increases under the ideal case, while it gets lowered when the interaction effect is included. For a weak interaction strength, the evolution of $\langle d\rangle$ agrees well with the ideal one for short evolution times, but becomes relatively smaller when propagating further. This should attribute to the effect from the trap potential and the initial momentum broadening. Additionally, the stepwise increase of $\langle d\rangle$, an inherent feature for the Floquet systems \cite{Goldman2014a}, is due to the interference of the multi-frequency Bragg laser beams, as discussed in Ref.~\cite{Gou2020}. As the interaction strength approaches the infinite limit (compared with the tunneling strength), $\langle d\rangle \to 0$. For a finite evolution time, as shown in Fig.~\ref{fig7}(e), we observe that $\langle d\rangle$ decreases as we change the scattering length from zero to $1300a_0$ (correspondingly, $\bar{U}_0/J$ changes from 0 to $3$ for parameters used there). For $\bar{U}_0/J < 0.5$, $\langle d\rangle$ slightly changes compared with the value under non-interacting case ($a_s=0$); while when $\bar{U}_0/J > 2$, the self-trapping significantly influences the dynamics, as indicated by Fig.~\ref{fig7}(c) where $\bar{U}_0/J = 2.25$. The difference between the value of $\langle d\rangle$ calculated with the GPE for zero $\bar{U}_0$ and that for an ideal tight-binding Hamiltonian is caused by both the initial momentum broadening and the interference of the multi-frequency Bragg laser components. 

\subsection{Topological phase transition in the SSH model under interactions}

Compared to the real-space optical lattice, the momentum-state lattice shows an outstanding capability for local controls on the tight-binding parameters \cite{Gadway2015}, which makes it significantly suitable to study the topological SSH model where the intra- and inter-cell tunnelings are set independently. The Hamiltonian for a general ($2M+1$)-cell SSH chain reads
\begin{eqnarray}\label{eq14}
 H &=& -\sum\limits_{m=-M}^{M} (J+\Delta_J) |2m-1\rangle\langle 2m| \nonumber\\
 & & -\sum\limits_{m=-M}^{M-1} (J-\Delta_J)|2m\rangle\langle 2m+1| + {\rm h.c.}, 
\end{eqnarray}
where $m$ is the cell index, and the $m$-th cell is comprised by the $|2m-1\rangle$ and $|2m\rangle$ sites. In the GPE calculations, we set the intra-cell tunneling $J_{\ell = 2m-1}= J + \Delta_J$ while the inter-cell tunneling $J_{\ell=2m} = J-\Delta_J$. As demonstrated in Refs.\cite{Cardano2017,Maffei2018}, the topological phase transition is monitored by the mean chiral displacement, $\mathcal{C}(t)=2\sum_m m (P_{2m-1}-P_{2m})$, which can be resolved by measuring the population dynamics in the chain. In previous experiment \cite{Xie2020}, we find that the time-averaged mean chiral displacement $\bar{\mathcal{C}}_A(t)$ converges to the value of winding number more rapidly than $\mathcal{C}(t)$. Here we use $\bar{\mathcal{C}}_A$ defined by 
\begin{equation}
 \bar{\mathcal{C}}_A(t) = \frac{1}{t}\int_0^t \mathcal{C}(t')dt'
\end{equation}
to describe the topological phase transition in our calculations. 

\begin{figure}[]
\includegraphics[width=0.5\textwidth]{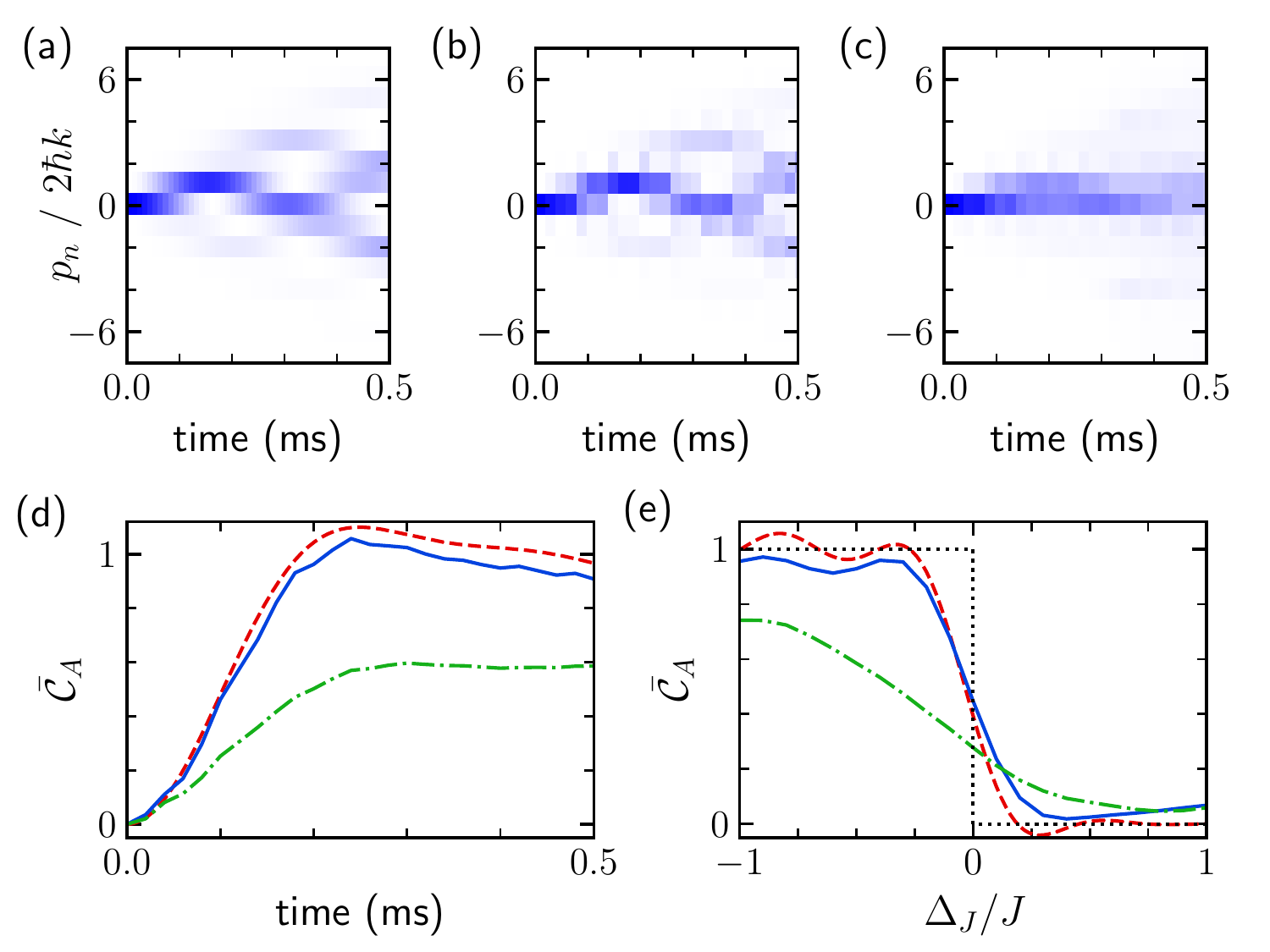}
\caption{\label{fig8} (Color online) Effect of many-body interactions on the topological phase transition for the SSH model. Here we consider a finite 14-site momentum-state lattice ($n\in [-7,6]$), i.e., 7 unit cells in total (the cell index $m \in [-3,3]$). Time evolutions of the populations in each site are calculated with: the tight-binding Hamiltonian (a), or the GPE model (\ref{eq3}) with $a_s=100 a_0$ (b) and $a_s=1000 a_0$ (c) during $t > 0$, respectively. Initially ($t<0$), the condensate ($N=6\times 10^4$, $a_s=100 a_0$) is prepared in a trap with frequencies $(\omega_x, \omega_y, \omega_z) = 2\pi \times (100, 100, 40)~{\rm Hz}$. In (a)-(c), we choose $J =2\pi \times 1000~{\rm Hz}$, and $\Delta_J = -2\pi\times 500~{\rm Hz}$ (see the maintext). (d) Evolutions of the time-averaged mean chiral displacement $\bar{\mathcal{C}}_A$. The red dashed, blue solid and green dotted dash lines respectively correspond to the population dynamics shown in (a)-(c). (e) Topological phase transition indicated by $\bar{\mathcal{C}}_A (t=0.5~{\rm ms})$ changing from 1 to 0 for a finite SSH chain under interactions. The red dashed (black dotted) line indicates the prediction from the tight-binding Hamiltonian with finite 7 unit cells (infinite length chain), while the blue solid (green dotted dash) line gives the result calculated from the GPE model (\ref{eq3}) with $a_s=100 a_0$ ($1000 a_0$). Here we fix $J=2\pi\times 1000~{\rm Hz}$ and vary $\Delta_J$ from $-J$ to $J$ to change the ratio between the intra- and inter-cell tunneling amplitudes. 
}
\end{figure}

To show the effect of the many-body interactions, we compare the dynamics of both the populations and $\bar{\mathcal{C}}_A$ calculated with the ideal tight-binding Hamiltonian and the GPE model respectively, as illustrated in Fig.~\ref{fig8}. Under weak interaction condition [e.g. $a_s=100a_0$ in Fig.~\ref{fig8}(b)], the time evolution of the population still shows similar behaviors to that obtained from the ideal tight-binding Hamiltonian (\ref{eq14}) [see Fig.~\ref{fig8}(a)], resulting in similar profiles of the dynamics of $\bar{\mathcal{C}}_A(t)$ [dashed and solid lines in Fig.~\ref{fig8}(d)]. When we enlarge $a_s=1000a_0$ to move into the strong interaction regime, a fraction of the atoms tend to be localized at the initial site $|0\rangle$, i.e., the self-trapping plays an essential role, as shown in Fig.~\ref{fig8}(c). So naturally, the resulted $\bar{\mathcal{C}}_A(t)$ [dotted dash line in Fig.~\ref{fig8}(d)] becomes relatively smaller than that under a weak interaction strength, as the populations in the central $m=0$ unit cell (site $|0\rangle$ and $|-1\rangle$ here) make no contribution to $\bar{\mathcal{C}}_A$. 

To check the influence of such interaction induced localizations on the topological phase transition, we plot $\bar{\mathcal{C}}_A$ at $t=0.5 ~{\rm ms}$ (it converges to nearly a constant value with the parameters used there) versus the ratio $\Delta_J/J$ in Fig.~\ref{fig8}(e). Ideally, for an SSH chain with an infinite length ($M\to\infty$), it lies in a topological phase ($\bar{\mathcal{C}}_A=1$) with a protected edge state when $\Delta_J/J<0$, while in a topologically trivial phase ($\bar{\mathcal{C}}_A=0$) when $\Delta_J/J>0$. For a finite-length chain ($M=3$), the phase transition becomes not as sharp as that for the ideal infinite case. Generally, now $\bar{\mathcal{C}}_A>0.5$ when $\Delta_J/J<0$ while $\bar{\mathcal{C}}_A<0.5$ when $\Delta_J/J>0$. However, things become different once the interaction terms are taken into account with the GPE model (\ref{eq3}). As shown in Fig.~\ref{fig8}(e), the transition under weak interaction strength (solid line) slightly differs from that calculated with the tight-binding Hamiltonian (\ref{eq14}), while the values of $\bar{\mathcal{C}}_A$ in $\Delta_J/J<0$ (topologically non-trivial) regime get significantly lowered when we enlarge the interaction strength. This indicates that a sufficiently strong interaction strength can make the topological phase transition collapse, as the inherent chiral symmetry of the system gets broken when the interaction terms are added to the Hamiltonian (\ref{eq14}).

\section{conclusion and outlook}\label{sec5}

In summary, we have developed a GPE description to the synthetic momentum-state lattice, which naturally and exactly addresses the trap potential and the mean-field many-body interactions. Such GPE models, either with the approximated decomposition applied or not, enable a more precise interpretation to the experimental observations, in comparison with the tight-binding Hamiltonian. Also, the non-resonant coupling terms are included by directly propagating the GPE (\ref{eq3}) followed by a Fourier transformation to the spatial wavefunction.

The roles of the trap and the interactions are clearly presented, by making detailed investigations on both the two-site oscillation and the multiple-site transport. In the studies of the two-site case, we find that, both the oscillation period and the decoherence time constant are significantly modified when either the trap potential or the interaction term is introduced. A relatively shallow trap (especially in the direction along which the Bragg laser is applied) and a weak interaction strength will make the oscillation approach the ideal Rabi oscillation predicted by the tight-binding model. In the multiple-state transport, we focus on the effect of the self-trapping, where the $s$-wave scattering length is assumed to be arbitrarily varied to change the interaction strength in our calculation. In the strongly interacting regime, the condensate tends to be localized and the transport dynamics become impeded. The strong interaction strength also wrecks the signatures of the topologically non-trivial phase in an interacting SSH model due to the broken chiral symmetry. 

The GPE models presented here provide an accurate treatment to the interactions, thus allow extending the momentum-state lattice to many-body settings. Currently, the experiments based on the momentum-state lattice typically work in the weakly interacting regime (the $s$-wave scattering length for $^{87}{\rm Rb}$ is $\sim 100 a_0$), so the transport dynamics and the topological phase transition get slightly modified. As strong interactions lead to the occurrence of localizations, it will be interesting to study the interplay with the disorder and the topology in future experiments \cite{Wang2020, Yoshihito2019,Sylvain2019}, for example, for other atomic species with relatively easier tunability of the interactions. Additionally, once the interaction can be periodically tuned to go beyond the mean-field approximation (the Bogoliubov excitation becomes important) \cite{Clark2017,Chen2018,Wu2019,Nguyen2019}, the dynamics of the pair correlation and the thermalization process in momentum-state lattices will also be an interesting and attractive question. 

\begin{acknowledgements}

We acknowledge the support from the National Key R\&D Program of China under Grant No. 2018YFA0307200, the National Natural Science Foundation of China under Grant No. 12074337, Natural Science Foundation of Zhejiang province under Grant No. LZ18A040001, Zhejiang Province Plan for Science and technology No. 2020C01019 and the Fundamental Research Funds for the Central Universities.

\end{acknowledgements}

\bibliographystyle{apsrev4-2}
\bibliography{GPEformsl.bib}

\end{document}